\documentclass[preprint,12pt, a4paper]{elsarticle}

\usepackage{amssymb}
\usepackage{hyperref}

\usepackage{amsmath,amssymb,amsfonts}
\usepackage{algorithmic}
\usepackage{graphicx}
\usepackage{textcomp}
\usepackage{xcolor}
\def\BibTeX{{\rm B\kern-.05em{\sc i\kern-.025em b}\kern-.08em
    T\kern-.1667em\lower.7ex\hbox{E}\kern-.125emX}}

\usepackage{url}

\usepackage{hyperref}
\usepackage{balance}
\usepackage{graphicx}
\usepackage{tabularx}
\usepackage{booktabs}
\usepackage{enumerate}
\usepackage{bmpsize}
\usepackage{multirow}
\usepackage{xspace}
\usepackage{listings}
\usepackage{xcolor}
\usepackage{lipsum}
\usepackage{float}
\usepackage{listings}
\usepackage{comment}
\usepackage{paralist}
\usepackage{color}
\usepackage{wasysym}

\newcommand\update[1]{\textcolor{black}{{#1}}}
\newcommand\revision[1]{\textcolor{black}{{#1}}}

\newcommand{\major}[1]{\textcolor{black}{{#1}}}

\usepackage{paralist}
\usepackage{color}

\newcommand{\BNG}{BeamNG.tech\xspace}
\newcommand{\BNGPY}{BeamNGpy\xspace}

\newcommand{\nb}[2]{}

\newcommand{\framework}{SDC-Scissor\xspace}
\newcommand{\frameworklong}{\textbf{SDC-Scissor} (\textbf{SDC} co\textbf{S}t-effe\textbf{C}t\textbf{I}ve te\textbf{S}t \textbf{S}elect\textbf{OR})\xspace}

\definecolor{codegreen}{rgb}{0,0.6,0}
\definecolor{codegray}{rgb}{0.5,0.5,0.5}
\definecolor{codepurple}{rgb}{0.58,0,0.82}
\definecolor{backcolour}{rgb}{0.95,0.95,0.92}

\lstdefinestyle{mystyle}{
    backgroundcolor=\color{backcolour},   
    commentstyle=\color{codegreen},
    keywordstyle=\color{magenta},
    numberstyle=\tiny\color{codegray},
    stringstyle=\color{codepurple},
    basicstyle=\ttfamily\footnotesize,
    breakatwhitespace=false,         
    breaklines=true,                 
    captionpos=b,                    
    keepspaces=true,                 
    numbers=left,                    
    numbersep=5pt,                  
    showspaces=false,                
    showstringspaces=false,
    showtabs=false,                  
    tabsize=2
}

\journal{Science of Computer Programming}

\begin{document}

\begin{frontmatter}



\title{Cost-effective Simulation-based Test Selection in Self-driving Cars Software}


\author[zurich]{Christian Birchler}
\author[zurich]{Nicolas Ganz}
\author[lugano,zurich]{Sajad Khatiri}
\author[passau]{Alessio Gambi}
\author[zurich]{Sebastiano Panichella}

\address[zurich]{Zurich University of Applied Sciences, Switzerland}
\address[lugano]{Software Institute - USI Lugano, Switzerland}
\address[passau]{University of Passau, Germany}

\begin{abstract}
\revision{Simulation environments are essential for the continuous development of complex cyber-physical systems such as self-driving cars (SDCs). Previous results on simulation-based testing for SDCs have shown that many automatically generated tests do not strongly contribute to identification of SDC faults, hence do not contribute towards increasing the quality of SDCs.
Because running such ``uninformative'' tests generally leads to a waste of computational resources and a drastic increase in the testing cost of SDCs, testers should avoid them. However, identifying ``uninformative'' tests \emph{before} running them remains an open challenge.
Hence, this paper proposes SDC-Scissor, a framework that leverages Machine Learning (ML) to identify SDC tests that are unlikely to detect faults in the SDC software under test, thus enabling testers to skip their execution
and drastically increase the cost-effectiveness of simulation-based testing of SDCs software. 
Our evaluation concerning the usage of six ML models on two large datasets characterized by 22’652 tests showed that \major{SDC-Scissor achieved a classification F1-score up to 96\%.}
Moreover, our results show that \major{SDC-Scissor outperformed a randomized baseline in identifying more failing tests per time unit.} 
}

\noindent Webpage \& Video: \url{https://github.com/ChristianBirchler/sdc-scissor}

\end{abstract}


\end{frontmatter}


\section*{Metadata}
\label{sec:mdetadata}
\begin{table}[H]
{
\scriptsize
\caption{Code metadata (mandatory)}
\label{tab:metadata}
\begin{tabular}{|l|p{4.5cm}|p{5.5cm}|}
\hline
\textbf{Nr.} & \textbf{Code metadata description} & \textbf{Please fill in this column} \\
\hline
C1 & Current code version & \major{v2.1.0} \\
\hline
C2 & Permanent link to code/repository used for this code version &  \url{https://github.com/ChristianBirchler/sdc-scissor} \\
\hline
C3 & Legal Code License   & GNU General Public License (GPLv3) \\
\hline
C4 & Code versioning system used & Git \\
\hline
C5 & Software code languages, tools, and services used & Python 3.9, BeamNG.tech v0.24.0.2 \\
\hline
C6 & Compilation requirements, operating environments and dependencies & Windows 10 \\
\hline
C7 & If available, link to developer documentation/manual & \major{\url{https://sdc-scissor.readthedocs.io/en/latest/}} \\
\hline
C8 & Support email for questions &  birchler.chr@gmail.com, spanichella@gmail.com\\
\hline
\end{tabular}
}
\end{table}

\section{Introduction}
\label{sec:intro}
\revision{Cyber-physical systems (CPSs) are complex systems that leverage physical capabilities from hardware components~\cite{baheti2011cyber} and find applications in various domains \update{including Robotics, Transportation and Healthcare}. For instance, in the automotive domain, self-driving cars (SDCs) are one emerging example of CPS,  expected to impact the transport system of our society profoundly. 
Specifically, human driving errors cause more than 90\% of car accidents \cite{KalraPaddock:2016} and SDCs have the potential to reduce such errors and eliminate most of these accidents.}
However, the recent fatal crashes involving SDCs suggest that the advertised large-scale adoption of SDCs appears optimistic~\cite{baheti2011cyber}.

\revision{Automated testing of SDCs (and in general CPS) to ensure their proper behaviour is still an open research challenge~\cite{afzal2020study}.
We argue that enabling cost-effective testing automation in Continuous Integration (CI) pipelines for SDCs is critical to address 
the safety and reliability requirements of SDCs~\cite{KalraPaddock:2016,Kim2019}.} However, current SDC testing practices have several limitations: 
\begin{inparaenum}[(i)]
\item difficulty in testing SDCs using representative, safety-critical tests~\cite{Ingrand19};
\item difficulty in assessing SDC's behavior in different environmental conditions~\cite{KalraPaddock:2016}.
\end{inparaenum}

\revision{To deal with such safety-related challenges, there is an increasing interest in adopting agile development paradigms within the CPS safety-critical domains~\cite{ZampettiKPP22,abs-2107-09612} to identify hazards and elicit safety requirements iteratively~\cite{DBLP:conf/re/Cleland-HuangV18}. 
Consequently, researchers proposed the usage of Digital-Twins\footnote{A digital twin is a virtual representation of a real-time digital counterpart of a physical object or process.} technologies to simulate and test CPSs in a diversified set of scenarios  \cite{HuangSLFB21,BojarczukGLDH0S21,PiazzoniCAYSV21,SDCScissor,NguyenHG21} to support testing automation \cite{AlconTAC21,Wotawa21}, regression testing \cite{SDCScissor,abs-2107-09614}, 
and debugging \cite{SmithR21,RoyHACC21} activities. 
In this context, simulation-based testing has been suggested as a promising direction to improve the SDC testing practices~\cite{afzal2021simulation,afzal2018crashing,wang2021exploratory} because simulation environments enable efficient test execution, reproducible results, and testing under critical conditions~\cite{Gambi2019Police}.} 
Additionally, simulation-based testing can be as effective as traditional field operational testing~\cite{afzal2020study,DosovitskiyRCLK17}.
However, the testing space of simulation environments is infinite, which poses the challenge of exercising the SDC behaviors adequately~\cite{Gambi2019,abdessalem2018testing}.
Given the limited budget devoted to testing activities, it is paramount that developers test SDCs in a cost-effective fashion:
using test suites optimized to reduce testing effort (time) without affecting their ability to identify faults~\cite{Yoo:2010, DBLP:journals/tse/NucciPZL20,abdessalem2018testing}.

To increase SDC testing cost-effectiveness, we propose \frameworklong, a framework that leverages Machine Learning (ML) approaches 
for identifying tests that are unlikely to detect faults 
and skips them before their execution,
hence, reducing the time spent in executing tests. 
Specifically, we refer to tests that do not expose a fault as \emph{safe} and deem them irrelevant. On the contrary, we consider tests that expose a fault (e.g., an SDC drives out of the road) as relevant and refer to them as \emph{unsafe}.

\revision{
\framework exploits six ML models trained on SDC simulation-based tests features that can be computed before the actual test execution (i.e., input features) 
to classify whether the tests are safe or unsafe~\cite{SDCScissor,abs-2107-09614}.}

\revision{We originally proposed employing Machine Learning to classify simulation-based tests and select them in~\cite{SDCScissor} for making more cost-effective the testing of SDCs. This paper extends our original work by making the following contributions:}

\begin{itemize}

\item \revision{A structural refactoring and extension of SDC-Scissor framework to provide an 
extendable open API (e.g., facilitating the integration of other SDC simulation environments, or an interface to implement an own AI or an own test generator) as well as the  possibility of using the \textit{z} coordinate (defining a road position in a three-dimensional space), which increases the level of realism of generated tests (given the non-flat roads).}

\item \revision{ An extension of original datasets that include new configurations of the test subject (i.e., risk factors RF1, RF1.5 and RF2) and additional 14'107 simulations-based tests.} 

\item  \revision{We extended the automated, ML-based approach integrating more ML models trained on features of SDC simulation-based tests 
to classify whether SDC tests are safe or unsafe (computed before the actual test execution); }
\item \revision{ An empirical study comparing the cost-effectiveness of the proposed approach with a randomized baseline as well as a Mean Decrease in Gini analysis to describe the most important SDC features used 
by the ML models in identify unsafe tests.}

\item \revision{To enable future studies, we made SDC-Scissor compatible with the recent version of BeamNG.tech (BeamNG.tech v0.24.0.2), which allows the generation of more diverse tests, with the possibility to test multiple cars simultaneously.}

\end{itemize}

 \revision{Through a large empirical study concerning the usage of six ML models on two large datasets characterized by around 23'000 SDC simulation-based tests, we assessed the performance of \framework in optimizing simulation-based testing. Our evaluation 
 showed that SDC-Scissor achieved a higher classification F1-score (between 56\% and 96\%) with the best performing ML models and outperformed a randomized baseline in identifying failing tests as well as in reducing the time spent running uninformative (i.e., safe) tests.
 }

\section{\update{The \framework Tool}}
\label{sec:approach}

In this section, we give an overview of \framework's software architecture and its main usage scenarios (Fig.~\ref{fig:architecture}); \revision{we describe the simulation environment it uses (i.e., \BNG) and its main APIs (Fig.~\ref{fig:APIs}); finally, we discuss in details the components, the approach} and the technologies behind \framework.

\subsection{\framework Architecture Overview \& Main Scenarios}
\label{sec:overview}

\framework supports two main usage scenarios: \textit{Benchmarking} and \textit{Prediction}.
In the \textit{Benchmarking} scenario, developers leverage \framework to determine the best ML model(s) to classify SDC simulation-based tests as safe or unsafe.
In the \textit{Prediction} scenario, instead, developers use those model(s) to classify and select newly generated test cases.

\begin{figure*}[t!]
\centering 
	\includegraphics[width=0.99\textwidth]{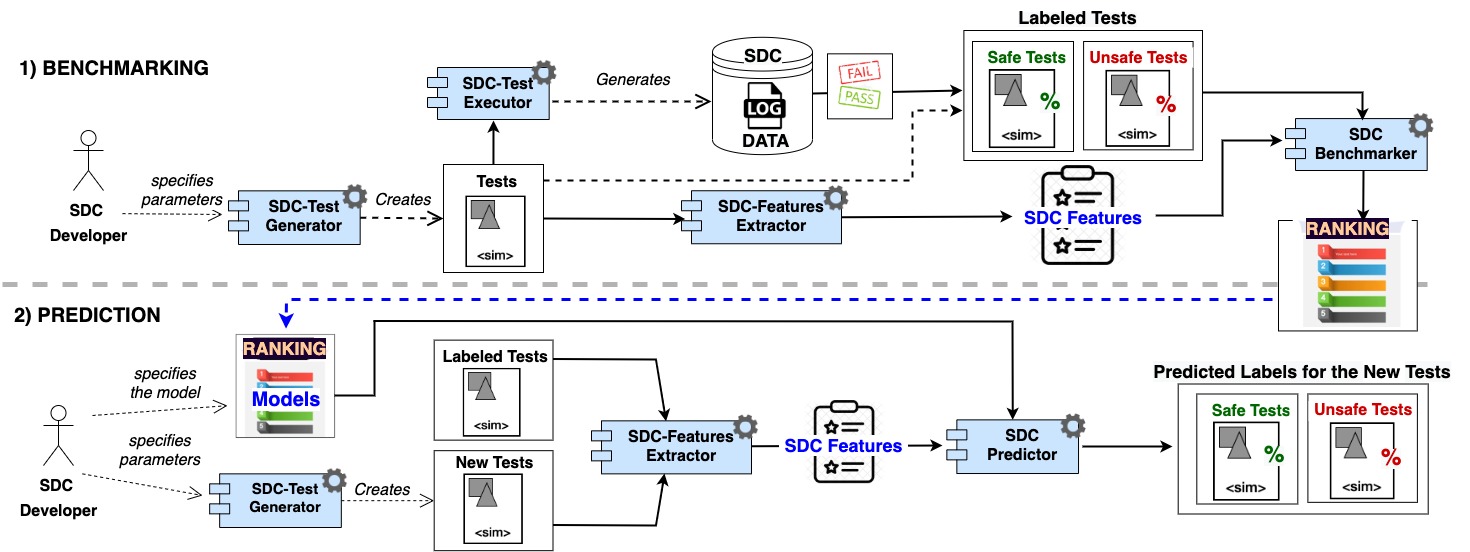} 
		\vspace{-2mm}
	\caption{The \framework's architecture.}
	\label{fig:architecture}
	\vspace{-5mm}
\end{figure*}

\framework Software Architecture implements these scenarios by means of the following software components (Fig.~\ref{fig:architecture}):
(i) \texttt{SDC-Test Generator} generates random SDC simulation-based tests, and (ii) \texttt{SDC-Test Executor} executes them. 
The test results produced by \texttt{SDC-Test Executor} are recorded and used to label tests as safe or unsafe; 
(iii) \texttt{SDC-Features Extractor} extracts input features of the executed SDC tests, while
(iv) \texttt{SDC-Benchmarker} uses these features and corresponding labels as input to train the ML models and determine which model best predicts the tests that are more likely to detect faults in SDCs;  
finally, (v) \texttt{SDC-Predictor} uses the ML models to classify newly generated test cases and enables test selection.

\subsection{\BNG's Simulation Environment}
\label{sec:swarchitecture}

\framework uses \BNG
to execute SDC tests as physically accurate and photo-realistic driving simulations.
\BNG can procedurally generate tests~\cite{Gambi2019} and was recently adopted in the ninth edition of the Search-Based Software Testing (SBST) CPS testing tool competition~\cite{SBST2021}. 

\BNG is organized around a central \emph{game engine} that communicates with the \emph{physics simulation}, the \emph{UI}, and the \emph{\BNGPY API}\footnote{\texttt{beamngpy} is available on PyPI and Github (\url{https://github.com/BeamNG/BeamNGpy})}. 
%
%
The UI can be used for game control and manual content creation (e.g., \emph{assets}, \emph{scenarios}).
For example, developers can use the world editor to create or modify the virtual environments that are used in the simulations; testers, instead, can create test scripts implementing driving scenarios (i.e., the tests).
The API, instead, allows the automated generation and execution of tests, the collection of simulation data (e.g., camera images, LIDAR point clouds) for training, testing, and validating SDCs. It also enables driving agents to drive simulated vehicles and get programmatic control over running simulations (e.g., pause/resume simulations, move objects around).
The \emph{game engine} manages the simulation setup, camera, graphics, sounds, gameplay, and overall resource management. The \emph{physics core}, instead, handles resource-intensive tasks such as collision detection and basic physics simulation; it also orchestrates the concurrent execution of the vehicle simulators. The \emph{vehicle simulators} ---one for each of the simulated vehicles--- simulate the high-level driving functions and the vehicle sub-systems (e.g., drivetrain, ABS).

We employ the BeamNG.AI\footnote{\url{https://wiki.beamng.com/Enabling\_AI\_Controlled\_Vehicles\#AI\_Modes}} lane-keeping system as the test subject for our evaluation: the driving agent is shipped with \BNG and drives the car by computing an ideal driving trajectory to stay in the center of the lane while driving within a configurable speed limit. 
As explained by \BNG developers, the \textit{risk factor} (RF) is a parameter that controls the driving style of BeamNG.AI: low-risk values (e.g., 0.7) result in smooth driving, whereas high-risk values (e.g., 1.7 and above) result in an edgy driving that may lead the ego-car to cut corners~\cite{SDCScissor}.

\subsection{The \framework's Approach and Technology Overview}
\label{sec:approach}
\framework integrates the extensible testing pipeline defined by the SBST tool competition\footnote{\url{https://github.com/se2p/tool-competition-av}} in its \texttt{SDC-Test Executor}.
We use the SBST tool competition infrastructure since it allows to (i) seamlessly execute the tests in  \BNG and (ii) distinguish between \textit{safe} and \textit{unsafe} tests based on whether the self-driving car keeps its lane (non-faulty tests) or depart from it (faulty tests)~\cite{Gambi2019}.
Consequently, \framework can accommodate various \texttt{SDC-Test Generators} for generating SDC simulation-based tests. In this paper, we demonstrate \framework by using the Frenetic test generation~\cite{CastellanoCTKZA21}, one of the most effective
tool submitted to the SBST tool competition.


\framework predicts whether the tests are likely to be safe or unsafe before their execution using input features that \texttt{SDC-Features Extractor} extracted. Specifically, this component extracts \textit{Full Road Features} (FRFs), i.e., a set of SDC features that describe global characteristics of the tests. Those features include the main \emph{road attributes} (see Table~\ref{table:road_general_feat}) and \emph{road statistics} concerning the road composition (see Table~\ref{table:road_segment_stat_feat}). 
Road statistics are calculated in three steps: (i) extraction of the \emph{reference driving path} that the ego-car has to follow during the test execution (e.g., the road segments that the car needs to traverse to reach the target position); (ii) extraction of metrics available \update{for} each road segment (e.g., length of road segments); and (iii) computation of standard aggregation functions on the collected road segments metrics (e.g., minimum and maximum). 

\begin{table}[t]
\scriptsize
    \centering
    \caption{Full Road Attributes extracted by the \textit{SDC-Features Extractor}}
    \label{table:road_general_feat}
     \fontsize{8pt}{7pt}\selectfont
\major{
     \begin{tabular}{l l l}
     \toprule
       \textbf{Feature} & \textbf{Description} & \textbf{Range}\\
     \midrule
     direct\_distance & Euclidean dist. between start and end (m) & [0\hfill--\hfill490] \\
     road\_distance & Tot. length of the driving path (m) & [50.6\hfill--\hfill3,317] \\
     num\_l\_turns & Nr. of left turns on the driving path & [0\hfill--\hfill18]\\
     num\_r\_turns & Nr. of right turns on the driving path & [0\hfill--\hfill17] \\
     num\_straights & Nr. of straight segments on the driving path & [0\hfill--\hfill11]   \\
     total\_angle & Cumulative turn angle on the driving path & [105\hfill--\hfill6,420] \\
     \bottomrule
    \end{tabular}
}
\end{table}

\begin{table}[t]
\centering
\scriptsize
\caption{Full Road Statistics extracted by the \textit{SDC-Features Extractor}}
\label{table:road_segment_stat_feat}
     \fontsize{8pt}{7pt}\selectfont
    \major{
     \begin{tabular}{l l l}
     \toprule
     \textbf{Feature}  & \textbf{Description} & \textbf{Range}\\
     \midrule
     median\_angle & Median turn angle on the driving path (DP) & [30\hfill--\hfill330]   \\
     std\_angle & Std. Deviation of turn angles on the DP & [0\hfill--\hfill150] \\
     max\_angle & Max. turn angle on the DP & [60\hfill--\hfill345] \\
     min\_angle & Min. turn angle on the DP & [15\hfill--\hfill285] \\
     mean\_angle & Average turn angle on the DP & [52.5\hfill--\hfill307.5] \\
     \midrule
     median\_pivot\_off & Median turn radius on the DP & [7\hfill--\hfill47] \\
     std\_pivot\_off & Std. Deviation of turn radius on the DP & [0\hfill--\hfill22.5] \\
     max\_pivot\_off & Max. turn radius on the DP & [7\hfill--\hfill47] \\
     min\_pivot\_off & Min. turn radius on the DP & [2\hfill--\hfill47]  \\
     mean\_pivot\_off & Average turn radius on the DP & [5.3\hfill--\hfill47] \\
     \bottomrule
    \end{tabular}
    }
\end{table}
\framework relies on the \texttt{SDC-Benchmarker} to determine the ML model that best classifies the SDC tests that are likely to detect faults. It follows an empirical approach to do so: given a set of labeled tests and corresponding input features, \texttt{SDC-Benchmarker} trains and evaluates an ensemble of standard ML models using the well-established \texttt{sklearn}\footnote{\url{https://scikit-learn.org/}} library. Next, it assesses ML models' quality using either 10-fold cross-validation or a testing dataset; and, finally, selects the best performing ML models according to Precision, Recall, and F1-score metrics~\cite{SDCScissor}.
Noticeably, \framework can use many different ML models; however, in this work, we consider Naive Bayes~\cite{Caruana06anempirical}, Logistic Regression \cite{ref1}, Random Forests~\cite{TinKamHo1998},\revision{ Gradient Boosting~\cite{ke2017lightgbm}, Support Vector Machine~\cite{suthaharan2016support}, and Decision Tree~\cite{safavian1991survey}}. We do so because these ML models have been successfully used for defect prediction or other classification problems in Software Engineering~\cite{PanichellaSGVCG15,SorboPASVCG16}.

Finally, the \texttt{SDC-Predictor} uses the ML models 
to predict the likelihood that newly generated SDC tests are safe or not. Specifically, developers have the possibility to select the ML models recommended by the \texttt{SDC-Benchmarker} (considered most accurate), or they can select other models of their choice. 

\revision{
\subsection{\framework's main APIs}
\label{sec:APIs}
}

\begin{figure*}[t!]
\centering 
	\includegraphics[width=0.85\textwidth]{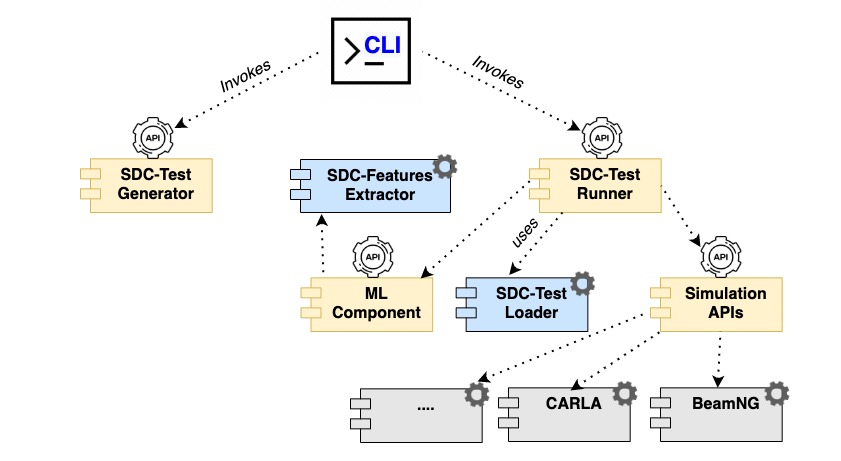} 
		\vspace{-2mm}
	 \revision{\caption{The \framework's main APIs.}}
	\label{fig:APIs}
	\vspace{-3mm}
\end{figure*}

\revision{
\framework was refactored and is now more modularized into components that offer APIs for enhancing better extensibility of the tool, as shown in Figure \ref{fig:APIs}.
The CLI component is where the user directly interacts with the tool, as described in Section \ref{sec:howTo}.
Furthermore, other test generators can be integrated by implementing the relevant API of the \texttt{SDC-Test Generator} component.
The main goal of the refactoring was to enable \framework to work with other simulators for the future (e.g., CARLA). For this purpose, we define \texttt{Simulation APIs} for simulators. The current version of \framework provides an implementation of the API for the BeamNG.tech simulator.
\framework also provides a \texttt{ML Component} and API for the training and testing the machine-learning models. This allows \framework to experiment easier on more diverse test selection approaches for the research on simulation-based regression testing on SDCs.\\
}

\section{Using \framework}
\label{sec:howTo}


\major{
\framework tool is openly available 
and can be used as a Python command-line utility via
\texttt{poetry}\footnote{\url{https://python-poetry.org/}} or \texttt{pip}. In the following sections it will be explained how \framework can be installed and used for \textit{Benchmarking} and \textit{Prediction} as shown in Figure~\ref{fig:architecture}.
}

\major{
\subsection{Installation}
}
\begin{lstlisting}[numbers=none]
git clone https://github.com/ChristianBirchler/sdc-scissor.git
cd sdc-scissor
poetry install
poetry run sdc-scissor [COMMAND] [OPTIONS]
\end{lstlisting}

To simplify \framework's usage, we also enable to execute it as a Docker\footnote{\url{https://www.docker.com}} container:
\begin{lstlisting}[numbers=none]
docker build --tag sdc-scissor .
docker run --volume "$(pwd)/results:/out" --rm
    sdc-scissor [COMMAND] [OPTIONS]
\end{lstlisting}

\begin{figure*}[t!]
\centering 
	\includegraphics[width=0.99\textwidth]{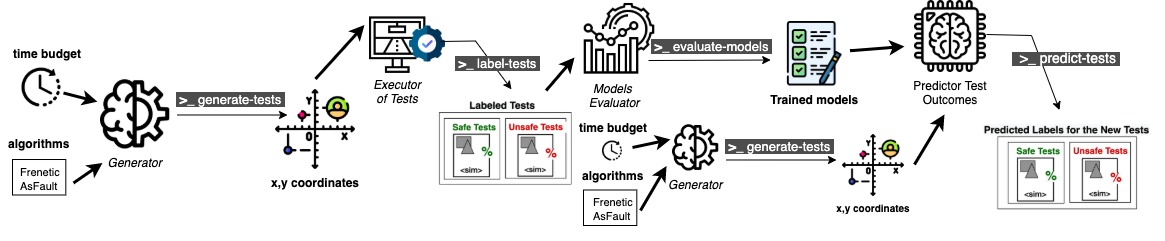} 
		\vspace{-4mm}
	\caption{The \framework's fine-grained view.}
	\label{fig:sdc-scissor}
	\vspace{-4mm}
\end{figure*}

As we detail below, \framework's command-line supports the execution of the main usage scenarios described in Section~\ref{sec:swarchitecture} by taking appropriate commands and inputs (see Fig.~\ref{fig:sdc-scissor}).

\major{
\subsection{Benchmarking}\label{sec:benchmarking}
\textbf{Test generation.} To generate SDC tests by running the Frenetic generator within a given number of desired tests, \framework requires the following command:
}
\begin{lstlisting}[numbers=none]
poetry run sdc-scissor generate-tests -c {number of tests to generate}
\end{lstlisting}

\textbf{Automated test labeling.}
\framework labels tests as safe and unsafe by executing them in BeamNG.tech.
Since BeamNG.tech cannot be run as a Docker container, labeling tests can be only run locally (i.e., outside Docker).
This labeling facility allows developers to create datasets that can be used for the training and validation of ML models (e.g., ML-based prediction of unsafe tests) \major{in the context of \textit{Benchmarking}}.
Generating a labeled dataset, requires a set of already generated SDC tests and the execution of the following command:

\begin{lstlisting}[numbers=none]
poetry run sdc-scissor label-tests -t /path/to/tests --rf {risk factor} --oob {OOB criteria}
\end{lstlisting}

If the car drives out of the lane to a certain percentage, also referred as the out of bound (OOB) criteria, then the test is labeled as unsafe.
\major{Based on the arguments for the risk factor and OOB the tests will be labeled. With different values for those arguments the tests can be labeled differently and therefore also affect the ML-based predictions.}

\major{
\textbf{Feature extraction.} The ML models requires as inputs features as described in Table~\ref{table:road_general_feat} and Table~\ref{table:road_segment_stat_feat}.
\framework extracts those features from the tests and stores them in a separate CSV file with the following command:
}

\begin{lstlisting}[numbers=none]
poetry run sdc-scissor extract-features -t /path/to/tests
\end{lstlisting}

\textbf{ML models evaluation.} For identifying the models that \framework could use for the prediction, \framework implements a 10-fold cross-validation strategy on the input labeled dataset. The following command tells \framework to benchmark all the configured ML models:

\begin{lstlisting}[numbers=none]
poetry run sdc-scissor evaluate-models --csv /path/to/road_features.csv
\end{lstlisting}

\major{
\subsection{Prediction}
For the prediction use case scenario we generate new tests with the same command as in Section~\ref{sec:benchmarking}.
The goal is to predict the test outcome before executing them.
For this reason we generate new tests for which we do not know the oracle yet.
}

\textbf{Test outcome prediction.} \framework classifies unlabeled tests, i.e., it predicts their outcome, using a trained ML model with the following command:
\begin{lstlisting}[numbers=none]
poetry run sdc-scissor predict-tests -t /path/to/tests
\end{lstlisting}

\textbf{Random baseline evaluation.} \framework allows to select tests using a random strategy that provides a baseline evaluation with the following command:

\begin{lstlisting}[numbers=none]
poetry run sdc-scissor evaluate-cost-effectiveness -csv /path/to/road_features.csv 
\end{lstlisting}



\begin{table}[t]
    \scriptsize
    \centering
\caption{Datasets Summary}
\label{table:datasets}

\begin{tabular}{l l r r r}
\toprule
\textbf{Dataset} &
\textbf{Test}
 & & \textbf{Data Points}\\
  & \textbf{Subject} & 
  \textbf{Unsafe} & \textbf{Safe} & \textbf{Total} \\
\hline
 & BeamNG.AI cautious 
& 1'318 (28\%) &  3'397 (72\%) &  4'715 \\ 
\textit{Dataset 1}  & BeamNG.AI moderate 
& 1'502 (34\%) &  2'908 (66\%) &  4'410 \\ 

 & BeamNG.AI reckless 
& 1'680 (34\%) &  3'302 (66\%) &  4'982 \\ 
\midrule
 & BeamNG.AI cautious 
 & 312 (26\%) & 866 (74\%) & 1’178 \\     
\textit{Dataset 2} & BeamNG.AI moderate 
& 2'543 (45\%) & 3'095 (55\%)  & 5’638 \\  
 &  BeamNG.AI reckless 
 & 1'655 (96\%) & 74 (4\%)      & 1’729 \\ 
\midrule
  &  
  \textbf{Total} & 9'010 (40\%)  &  13'642 (60\%)  &  22'652 \\ 
\bottomrule
\end{tabular}
\end{table}

\section{\major{Evaluation and Threats to Validity}}
\label{sec:evaluation}

\major{\subsection{\major{Evaluation}}}
\major{In this paper, we seek to answer
the following research questions:\\\\
 \textit{To what extent is it possible to predict safe and unsafe SDC test cases? To what extent SDC-scissor is cost-effectiveness compared to a random baseline?}}
\\

\major{We are interested to investigate the extent to which predicting unsafe SDC test cases before executing them is possible in a practical sense (e.g., do we achieve a reasonable precision, recall and F-measure?). More important, we also investigate whether \framework allows to reduce testing cost dedicated to the execution of so called irrelevant tests, i.e., test cases not leading to actual faults. To achieve these objectives, as described below, we first of all constructed a dataset of SDC tests cases that can be used to experiment with such research questions. Hence, we specifically investigated the usage of SDC road features to predict SDC test outcomes as well as investigate the ability of \framework in outperforming a random baseline.  Finally, we also discuss the most important features used for enabling the prediction in the context of our work.}

\major{
\textbf{Dataset construction}. We evaluated \framework
conducting a large study on two datasets, referred as \textit{Dataset~1}  and \textit{Dataset~2}, that contain \revision{over \update{$22,000$} SDC tests (see Table~\ref{table:datasets})}.}
We adopted the following experimental setup to obtain comprehensive and unbiased training datasets.
For \textit{Dataset 1}, we \emph{randomly} generated \revision{$13,207$ valid tests using Frenetic ~\cite{CastellanoCTKZA21} as well as collected input features and executed them to collect labels}. 
For the \textit{Dataset 2}, instead, we generated \update{$8,545$} tests using AsFault \cite{Gambi2019}.

\major{\textbf{AI engine and risk factor considered}. It is important to note that in executing all those tests, we experimented with different BeamNG.AI's risk factor as it influences the ego-car driving style. 
Specifically, we considered three configurations: cautious (RF $1.0$), moderate (RF $1.5$), and reckless (RF $2.0$) driver. 
Using different values for the risk factor enabled us to study the effectiveness of \framework on various SDCs' driving styles. 
We empirically validated our expectations by running the cautious, moderate, and reckless drivers to generate both \textit{Dataset~1} and \textit{Dataset~2} tests.
From Table \ref{table:datasets} we can observe that the number of unsafe tests increased with increasing values of BeamNG.AI's risk factor. 
This result seems to suggest that the risk factor may influences the safety of BeamNG.AI and the outcome of tests.
However, it is important to notice that for \textit{Dataset 1}, the ratio of safe (66\%) and unsafe (34\%) tests between moderate (RF $1.5$) and reckless (RF $2.0$) drivers is identical.}

\begin{table}[t]
\scriptsize

\caption{Performance of the \major{best three} ML models with dataset split 80/20. The best results are shown in boldface.}
\label{tab:risk_factor}
\centering
\begin{tabular}{lllllllll}
\toprule
\textbf{Dataset} & \textbf{RF} & \textbf{Model} &\textbf{Prec.} & \textbf{Recall} & \textbf{F1-score} \\
\midrule
&& Logistic \major{Regression}  & \textbf{40.3\%} & 55.5\% & \textbf{46.7\%}  \\
\textit {Dataset 1} & \textbf{RF 1} & Naïve Bayes  & \textbf{40.3\%} & 49.8\% & 44.6\% \\
&& Random Forest  & 38.9\% & \textbf{57.5\%} & \textbf{46.4\%}  \\
\midrule
& & Logistic \major{Regression} & \textbf{45.8\%} & 60.9\% & 52.3\%  \\
\textit {Dataset 1} &  \textbf{RF 1.5}  & Naïve Bayes & 40.2\% & \textbf{92.5\%} & \textbf{56.1\%}  \\ 
& & Random Forest & 41.3\% & 30.5\% & 35.1\%  \\
\midrule
&& Logistic \major{Regression}  & \textbf{39.4\%} & 53.6\% & 45.5\%  \\
\textit {Dataset 1} & \textbf{RF 2} & Naïve Bayes  & 34.6\% & \textbf{100.0\%} & \textbf{51.4\%} \\
&& Random Forest  & 38.3\% & 53.3\% & 44.6\%  \\
\hline
\hline
& & Logistic \major{Regression}  & \textbf{43.3\%} & 87.3\% & \textbf{57.9\%} \\
\textit {Dataset 2} &  \textbf{RF 1}  & Naïve Bayes  & 36.7\% & \textbf{92.1\%} & 52.5\%  \\ 
& & Random Forest  & 40.7\% & 79.4\% & 53.8\%  \\
\midrule
&& Logistic \major{Regression}  & 78.1\% & \textbf{65.3\%} & \textbf{71.1\%}  \\
\textit {Dataset 2} & \textbf{RF 1.5} & Naïve Bayes  & \textbf{79.3\%} & 53.2\% & 63.6\% \\
&& Random Forest  & 75.8\% & 62.7\% & 68.6\%  \\
\midrule
&& Logistic \major{Regression}  & \textbf{99.6}\% & 82.8\% & 90.4\% \\
\textit {Dataset 2} & \textbf{RF 2}  & Naïve Bayes  & 98.7\% & \textbf{94.3\%} & \textbf{96.4\%}  \\
&& Random Forest & \textbf{99.7\%} & 92.7\% & \textbf{96.1\%}  \\
\bottomrule
\end{tabular}
            \vspace{-2mm}
\end{table}

\major{\textbf{ML models and training process considered}.
To assess the performance of \framework in optimizing simulation-based SDCs testing via test selection (i.e., in selecting unsafe tests before executing them), for both \textit{Dataset 1} and \textit{Dataset 2} we experimented with the ML models 
mentioned in Section~\ref{sec:approach} trained and validated using an 80/20 data split.}

\major{\textbf{Results}.
As reported in Table \ref{tab:risk_factor}, on \textit{Dataset 1} \framework accurately identified unsafe test cases, with F1-score ranging between \update{35.1\%} and \update{56.1\%}.
On \textit{Dataset 2}, instead, \framework identified unsafe test cases with F1-score ranging between \update{52.5\%} and \update{96.4\%}.}

\major{\textbf{Cost-effectiveness}.
In the context of regression testing we want to select only relevant test scenarios so that the testing cost (execution time) is reduced.
We evaluated the cost-effectiveness of SDC-Scissor by computing the ratio selected unsafe test scenarios and the overall test execution time.
Thus, the cost-effectiveness is computed as
\[CE = \frac{\text{number of selected unsafe tests}}{\text{simulation time of all selected tests}}.\]
We compared the cost-effectiveness of \framework with a random baseline test selector.
In case of \framework, the models were trained on 80\% of \textit{Dataset 1 RF 1.5}.
\framework selected from the remaining 20\% 10 tests that are most likely to be unsafe, whereas the random baseline selector picks 10 tests at random.
As shown in Table~\ref{tab:cost-effectiveness}, \framework has only in the case of the \textit{Decision Tree} model a worse cost-effectiveness of 2.3\permil against the baseline with a cost-effectiveness of 2.6\permil.
For the \textit{Gradient Boosting}, \textit{Support Vector Machine}, and \textit{Logistic Regression} classifiers we have the highest differences of more than 1\permil.
Overall, we observed a better cost-effectiveness of \framework compared to a random baseline test selector.
In general, with our approach we detect more unsafe tests as the baseline per time unit.
}

\major{
\begin{table}[]
    \centering
    \major{
    \caption{Cost-effectiveness ($=\frac{\text{number of unsafe tests selected}}{\text{simulation time of all selected tests}}$) of \framework against a random baseline on \textit{Dataset 1} with \textit{RF 1.5}.}
    \begin{tabularx}{\textwidth}{Xcc} \toprule
        \multirow{2}{*}{\textbf{Model}} & \multicolumn{2}{c}{\textbf{Cost-effectiveness}}   \\ \cline{2-3}
                                        & \textbf{\framework}   & \textbf{Random Baseline}  \\ \midrule
        Random Forest                   & 2.9\permil                & 2.2\permil                    \\
        Gradient Boosting               & 4.0\permil                & 1.9\permil                    \\
        SVM                             & 3.5\permil                & 1.9\permil                    \\
        Naive Bayes                     & 2.4\permil                & 2.2\permil                    \\
        Logistic Regression             & 3.7\permil                & 2.2\permil                    \\
        Decision Tree                   & 2.3\permil                & 2.6\permil                    \\ \bottomrule
    \end{tabularx}
    \label{tab:cost-effectiveness}
    }
\end{table}
}

\begin{figure}[ht]
    \centering
    \includegraphics[width=0.78\textwidth]{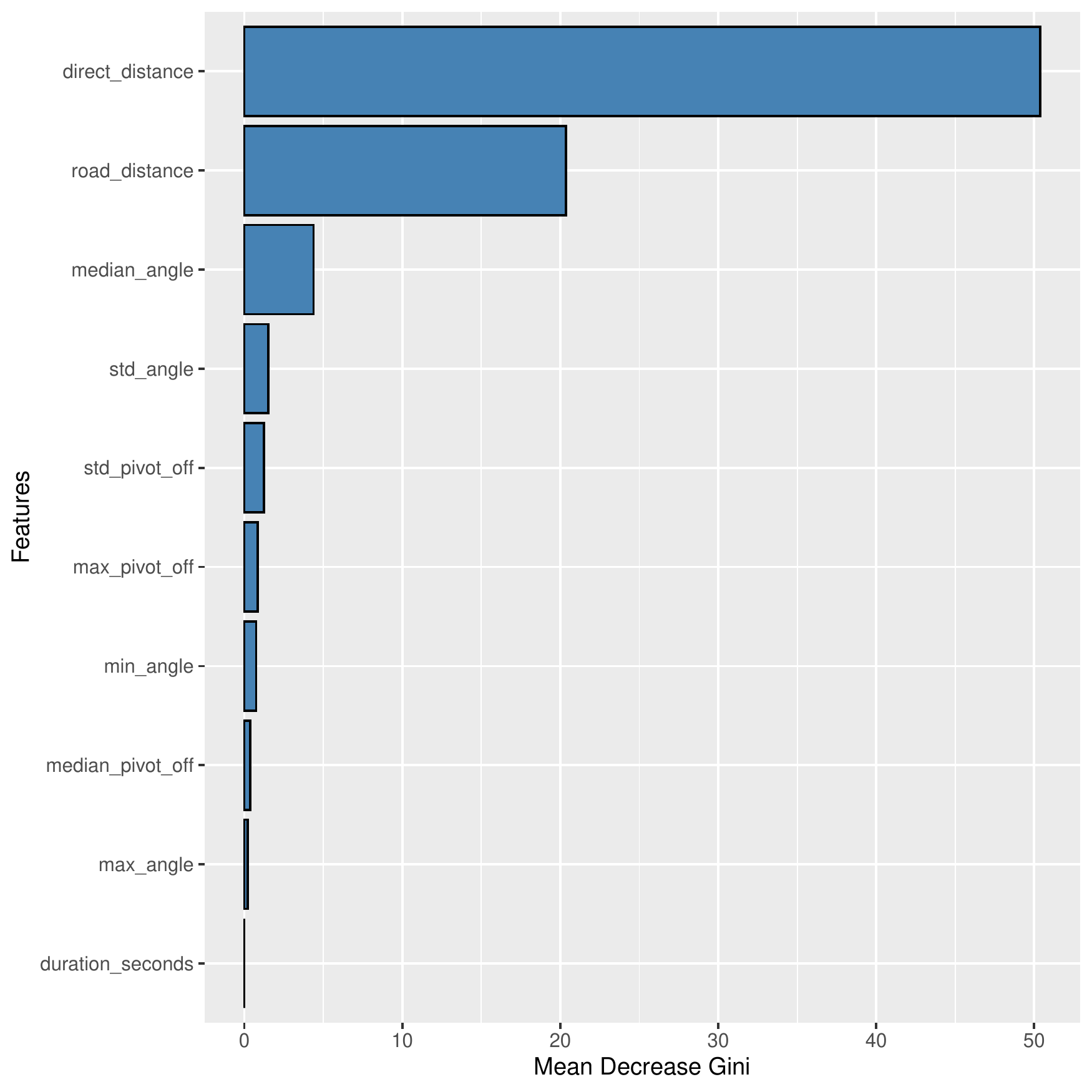}
    \caption{ 
	\revision{Mean Decrease in Gini when using RF 1.0. The top 10 features are visualized \major{(simulation time attributes included)}.}} 
	\label{fig:MDI-RF1}
\end{figure}

\begin{figure}[ht]
    \centering
    \includegraphics[width=0.78\textwidth]{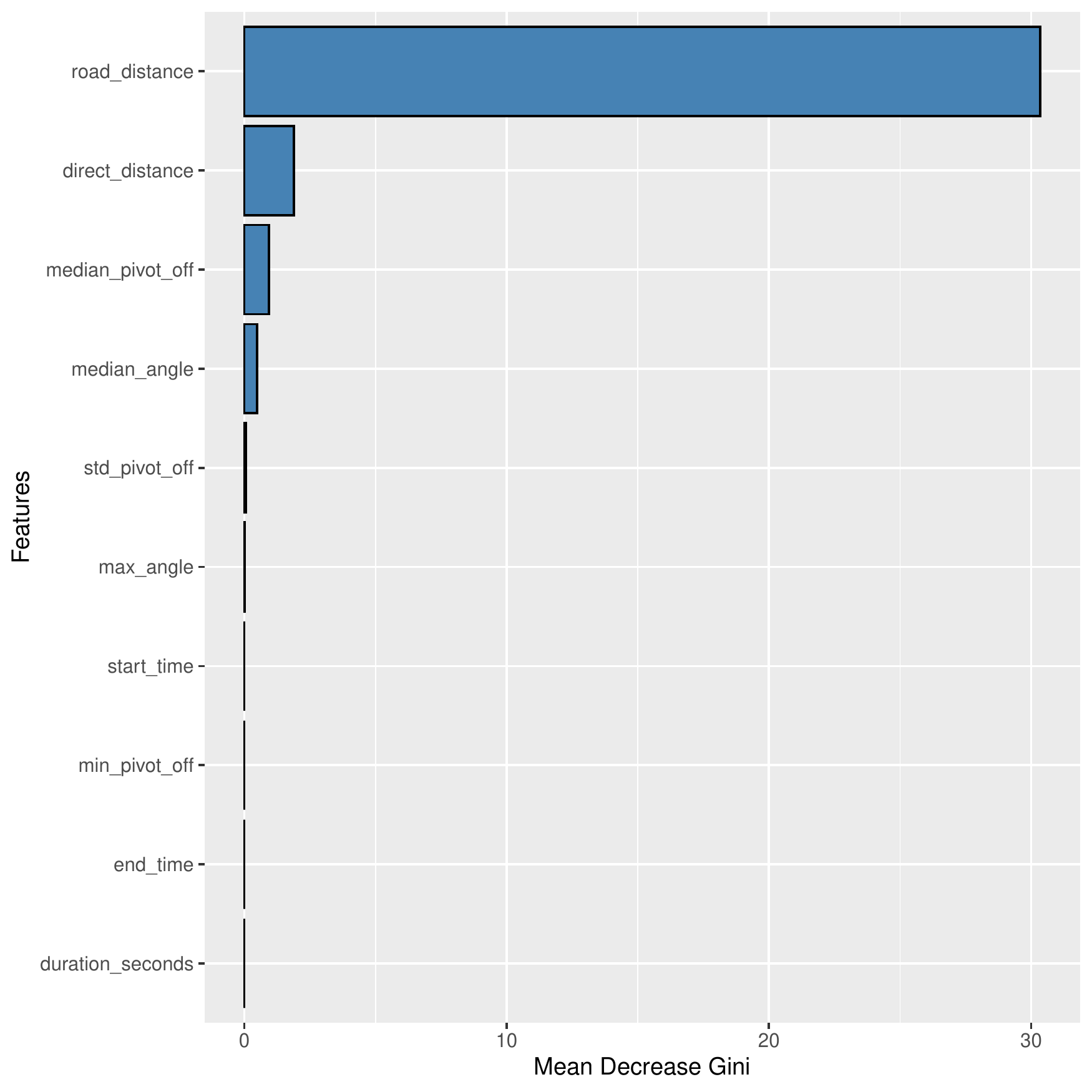}
    \caption{\revision{
	Mean Decrease in Gini when using RF 1.5. The top 10 features are visualized \major{(simulation time attributes included)}.}} 
	\label{fig:MDI-RF1-5}
\end{figure}

\begin{figure}[ht]
    \centering
    \includegraphics[width=0.78\textwidth]{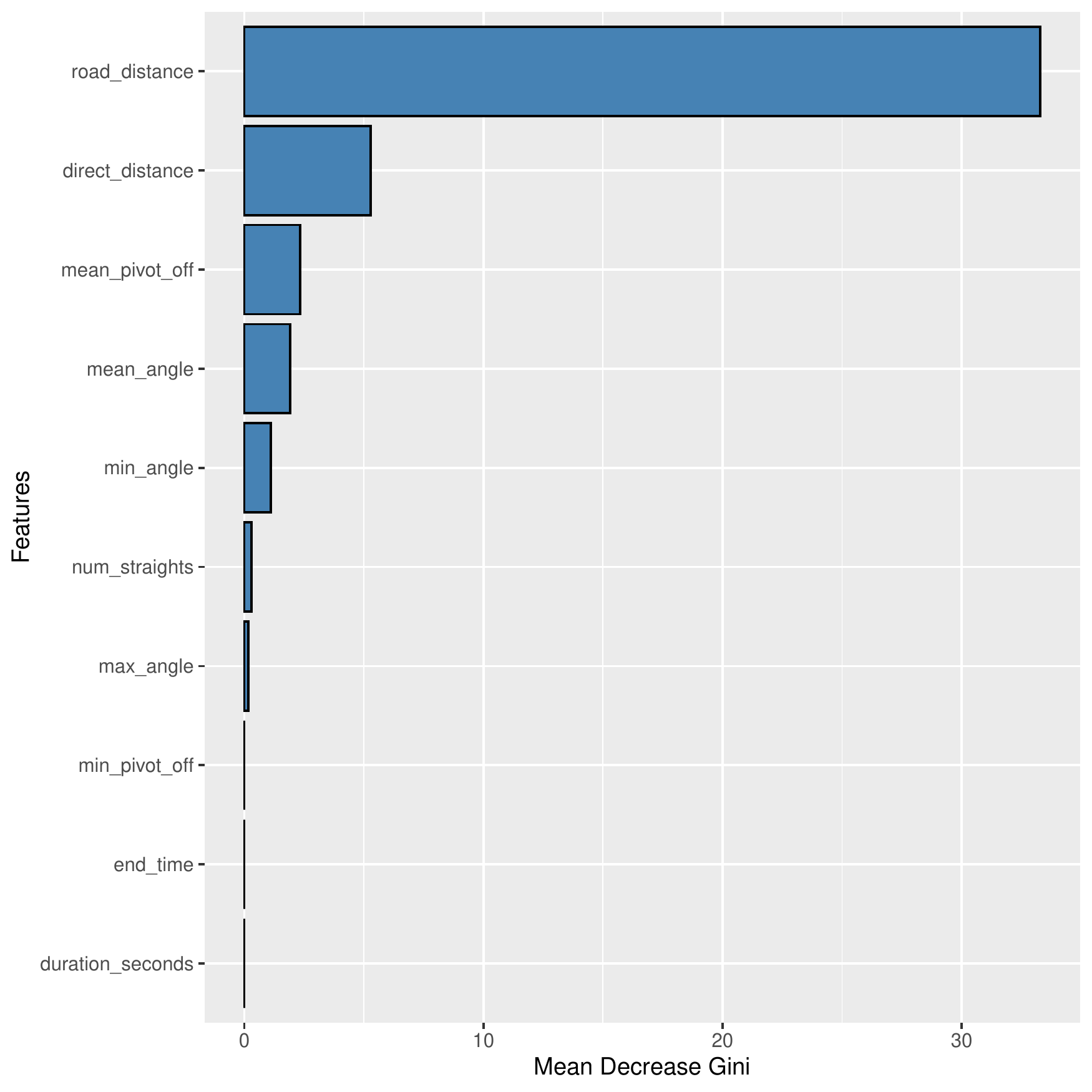}
    \caption{\revision{
	Mean Decrease in Gini when using RF 2.0. The top 10 features are visualized \major{(simulation time attributes included)}.}} 
	\label{fig:MDI-RF2}
\end{figure}

\revision{\textbf{Feature Importance}.
To better understand the features that contribute more to the prediction of safe and unsafe tests, we computed the Mean Decrease in Gini (also called Mean Decrease in Impurity) \cite{MDG:2020,GerstenbergerV15,9240701} 
considering the SDC designed road features.
%
As shown in Figure \ref{fig:MDI-RF1}, Figure \ref{fig:MDI-RF1-5} and Figure \ref{fig:MDI-RF2}, we can find the (top 10) features considered as important for the identification of safe and unsafe tests, for different risk factors (RF1.0, RF1.5 and RF2). 
%
%
It is interesting to observe from such features that the three top most important features vary depending on the specific configuration of the driving agent (i.e., RF). 
This observation suggests that certain characteristics of the road play an important role on the safety of the SDC, depending on the driving style (i.e., each RF). 
Specifically, for RF~1.0, the top thee most important road features are the \textit{Direct Distance},  \textit{Road Distance}, and \textit{Median Angle}. For RF~1.5, the top three most important road features are  the \textit{Road Distance},  \textit{Direct Distance}, and  \textit{Median Pivot Off}, while for RF~2.0, the top three most important road features are the \textit{Road Distance},  \textit{Direct Distance}, and  \textit{Mean Pivot Off}.
Hence, for less cautious driving styles (for RF $> 1.0$), the most important feature is always represented by the \textit{Road Distance}, followed by the \textit{Direct Distance} feature and the \textit{Mean/Median Pivot Off} feature. Finally, for more cautious driving style (for RF $=1.0$),  the most important feature is represented by the \textit{Direct Distance}, followed by the \textit{Road Distance} and the \textit{Median Angle} features.
In a practical sense, this means that for a more cautious driving style (for RF $= 1.0$) the safety of the SDC is influenced by the direct/road distance and the turn angle on the driving path (i.e., the distance and the presence of turns are together influencing the SDC behavior).
Complementary, for a less cautious driving style (for RF $> 1.0$) the safety of the SDC is influenced by the direct/road distance and the  average/median radius of road segment turned on the test track (i.e., the distance and the radius of specific road segments are together influencing the SDC behavior).
}

\major{\subsection{\major{ Threats to Validity}}
\framework is a ML-based test selector that depends on the data for training the models.
The datasets were labeled with the internal BeamNG.AI of the used BeamNG simulator.
The use of a single AI engine may introduce a threat to validity since the results might be biased since no other experiments with different AI were considered.
Furthermore, we do not know how the BeamNG.AI behaves with different weather conditions, which would increase the level of realism.
The use of different simulators with different physical behavior could alter the results because BeamNG is a soft-body physics simulator with high fidelity that simulates deformations of multiple parts of the car such as the chassis, engine, transmission, tires, etc. whereas other simulators like CARLA use a rigid-body physics engine.
Furthermore, the ML models are trained with the default configurations.
The prediction performances might be improved so that results changes and the ranking of the models vary.
}

\section{Conclusions}
\label{sec:conclusion}
%

This paper presented \framework, a ML-based test selection approach that classifies SDC simulation-based tests as likely (or unlikely) to expose faults before executing them. 
\framework trains ML models using input features extracted from driving scenarios, i.e., SDC tests, and uses them to classify SDC tests before their execution. Consequently, it selects only those tests that are predicted to likely expose faults. Our evaluation shows that \framework successfully selected unsafe test cases across different driving styles and drastically reduced 
the execution time dedicated to executing safe tests compared to a random baseline approach.

As future work, \revision{we plan to replicate our study on further SDC datasets, AI engines and more advanced SDC features to study how the results generalize in various autonomous systems domains}. 
Additionally, given our close contacts with the \BNG team, we plan the integration of \framework into \BNG environment to enable researchers and SDC developers to use \framework as a cost-effective testing environment for SDCs.
Finally, we plan to investigate the use of \framework in other CPS domains, such as drones, to investigate how it performs when testing focuses on different types of safety-critical faults. \revision{Specifically, it is important to investigate approaches that are more human-oriented or are able to integrate humans into-the-loop~\cite{PanichellaSGVCG15,SorboPASVCG16}, via multi-objective optimizations ~\cite{CanforaLPOPP13,GranoLPP21}}.

\major{
\framework can be used in an industrial context to identify relevant test scenarios. When it comes to different levels of testing like Software-in-the-loop or Hardware-in-the-loop, \framework provides a platform to conduct those experiments without manual human-based interaction. The testing costs can be reduced and the fault detection rate is increased compared to a random test selector.
}

\section*{Acknowledgements}
{
\small We gratefully acknowledge the Horizon 2020 (EU Commission) support for the project \textit{COSMOS} (DevOps for Complex Cyber-physical Systems), Project No. 957254-COSMOS) and the DFG project STUNT (DFG Grant Agreement n. FR 2955/4-1).
}

{
\scriptsize
\bibliographystyle{elsarticle-num}
\bibliography{main}

\begin{thebibliography}{10}
\expandafter\ifx\csname url\endcsname\relax
  \def\url#1{\texttt{#1}}\fi
\expandafter\ifx\csname urlprefix\endcsname\relax\def\urlprefix{URL }\fi
\expandafter\ifx\csname href\endcsname\relax
  \def\href#1#2{#2} \def\path#1{#1}\fi

\bibitem{baheti2011cyber}
R.~Baheti, H.~Gill, Cyber-physical systems, The impact of control technology
  12~(1) (2011) 161--166.

\bibitem{KalraPaddock:2016}
N.~Kalra, S.~Paddock, Driving to safety: How many miles of driving would it
  take to demonstrate autonomous vehicle reliability?, Transportation Research
  Part A: Policy and Practice 94 (2016) 182--193.
\newblock \href {https://doi.org/10.1016/j.tra.2016.09.010}
  {\path{doi:10.1016/j.tra.2016.09.010}}.

\bibitem{afzal2020study}
A.~Afzal, C.~Le~Goues, M.~Hilton, C.~S. Timperley, A study on challenges of
  testing robotic systems, in: 2020 IEEE 13th International Conference on
  Software Testing, Validation and Verification (ICST), IEEE, 2020, pp.
  96--107.

\bibitem{Kim2019}
J.~Kim, S.~Chon, J.~Park,
  \href{https://doi.org/10.1109/icstw.2019.00043}{Suggestion of testing method
  for industrial level cyber-physical system in complex environment}, in:
  International Conference on Software Testing, Verification and Validation
  Workshops, 2019.
\newblock \href {https://doi.org/10.1109/icstw.2019.00043}
  {\path{doi:10.1109/icstw.2019.00043}}.
\newline\urlprefix\url{https://doi.org/10.1109/icstw.2019.00043}

\bibitem{Ingrand19}
F.~Ingrand, Recent trends in formal validation and verification of autonomous
  robots software, in: International Conference on Robotic Computing, 2019, pp.
  321--328.

\bibitem{ZampettiKPP22}
F.~Zampetti, R.~Kapur, M.~D. Penta, S.~Panichella,
  \href{https://doi.org/10.1016/j.jss.2022.111425}{An empirical
  characterization of software bugs in open-source cyber-physical systems}, J.
  Syst. Softw. 192 (2022) 111425.
\newblock \href {https://doi.org/10.1016/j.jss.2022.111425}
  {\path{doi:10.1016/j.jss.2022.111425}}.
\newline\urlprefix\url{https://doi.org/10.1016/j.jss.2022.111425}

\bibitem{abs-2107-09612}
A.~D. Sorbo, F.~Zampetti, C.~A. Visaggio, M.~D. Penta, S.~Panichella, Automated
  identification and qualitative characterization of safety concerns reported
  in uav software platforms, ACM Transactions on Software Engineering and
  Methodology (TOSEM) (2022).

\bibitem{DBLP:conf/re/Cleland-HuangV18}
J.~Cleland{-}Huang, M.~Vierhauser,
  \href{https://doi.org/10.1109/RE.2018.00034}{Discovering, analyzing, and
  managing safety stories in agile projects}, in: 26th {IEEE} International
  Requirements Engineering Conference, {RE} 2018, Banff, AB, Canada, August
  20-24, 2018, 2018, pp. 262--273.
\newblock \href {https://doi.org/10.1109/RE.2018.00034}
  {\path{doi:10.1109/RE.2018.00034}}.
\newline\urlprefix\url{https://doi.org/10.1109/RE.2018.00034}

\bibitem{HuangSLFB21}
Z.~Huang, Y.~Shen, J.~Li, M.~Fey, C.~Brecher,
  \href{https://doi.org/10.3390/s21196340}{A survey on ai-driven digital twins
  in industry 4.0: Smart manufacturing and advanced robotics}, Sensors 21~(19)
  (2021) 6340.
\newblock \href {https://doi.org/10.3390/s21196340}
  {\path{doi:10.3390/s21196340}}.
\newline\urlprefix\url{https://doi.org/10.3390/s21196340}

\bibitem{BojarczukGLDH0S21}
K.~Bojarczuk, N.~Gucevska, S.~M.~M. Lucas, I.~Dvortsova, M.~Harman, E.~Meijer,
  S.~Sapora, J.~George, M.~Lomeli, R.~Rojas,
  \href{https://doi.org/10.1145/3475716.3484196}{Measurement challenges for
  cyber cyber digital twins: Experiences from the deployment of facebook's {WW}
  simulation system}, in: F.~Lanubile, M.~Kalinowski, M.~T. Baldassarre (Eds.),
  {ESEM} '21: {ACM} / {IEEE} International Symposium on Empirical Software
  Engineering and Measurement, Bari, Italy, October 11-15, 2021, {ACM}, 2021,
  pp. 2:1--2:10.
\newblock \href {https://doi.org/10.1145/3475716.3484196}
  {\path{doi:10.1145/3475716.3484196}}.
\newline\urlprefix\url{https://doi.org/10.1145/3475716.3484196}

\bibitem{PiazzoniCAYSV21}
A.~Piazzoni, J.~Cherian, M.~Azhar, J.~Y. Yap, J.~L.~W. Shung, R.~Vijay,
  \href{https://doi.org/10.1109/AITEST52744.2021.00035}{Vista: a framework for
  virtual scenario-based testing of autonomous vehicles}, in: 2021 {IEEE}
  International Conference on Artificial Intelligence Testing, AITest 2021,
  Oxford, United Kingdom, August 23-26, 2021, {IEEE}, 2021, pp. 143--150.
\newblock \href {https://doi.org/10.1109/AITEST52744.2021.00035}
  {\path{doi:10.1109/AITEST52744.2021.00035}}.
\newline\urlprefix\url{https://doi.org/10.1109/AITEST52744.2021.00035}

\bibitem{SDCScissor}
C.~Birchler, N.~Ganz, S.~Khatiri, A.~Gambi, S.~Panichella, Cost-effective
  simulation-based test selection in self-driving cars software with
  sdc-scissor, in: The 29th IEEE International Conference on Software Analysis,
  Evolution, and Reengineering, 2022.

\bibitem{NguyenHG21}
V.~Nguyen, S.~Huber, A.~Gambi,
  \href{https://doi.org/10.1109/AITEST52744.2021.00033}{{SALVO:} automated
  generation of diversified tests for self-driving cars from existing maps},
  in: 2021 {IEEE} International Conference on Artificial Intelligence Testing,
  AITest 2021, Oxford, United Kingdom, August 23-26, 2021, {IEEE}, 2021, pp.
  128--135.
\newblock \href {https://doi.org/10.1109/AITEST52744.2021.00033}
  {\path{doi:10.1109/AITEST52744.2021.00033}}.
\newline\urlprefix\url{https://doi.org/10.1109/AITEST52744.2021.00033}

\bibitem{AlconTAC21}
M.~Alcon, H.~Tabani, J.~Abella, F.~J. Cazorla,
  \href{https://doi.org/10.1109/DSD53832.2021.00071}{Enabling unit testing of
  already-integrated {AI} software systems: The case of apollo for autonomous
  driving}, in: F.~Leporati, S.~Vitabile, A.~Skavhaug (Eds.), 24th Euromicro
  Conference on Digital System Design, {DSD} 2021, Palermo, Spain, September
  1-3, 2021, {IEEE}, 2021, pp. 426--433.
\newblock \href {https://doi.org/10.1109/DSD53832.2021.00071}
  {\path{doi:10.1109/DSD53832.2021.00071}}.
\newline\urlprefix\url{https://doi.org/10.1109/DSD53832.2021.00071}

\bibitem{Wotawa21}
F.~Wotawa, \href{http://ceur-ws.org/Vol-2808/Paper\_29.pdf}{On the use of
  available testing methods for verification {\&} validation of ai-based
  software and systems}, in: H.~Espinoza, J.~McDermid, X.~Huang,
  M.~Castillo{-}Effen, X.~C. Chen, J.~Hern{\'{a}}ndez{-}Orallo, S.~{\'{O}}.
  h{\'{E}}igeartaigh, R.~Mallah (Eds.), Proceedings of the Workshop on
  Artificial Intelligence Safety 2021 (SafeAI 2021) co-located with the
  Thirty-Fifth {AAAI} Conference on Artificial Intelligence {(AAAI} 2021),
  Virtual, February 8, 2021, Vol. 2808 of {CEUR} Workshop Proceedings,
  CEUR-WS.org, 2021.
\newline\urlprefix\url{http://ceur-ws.org/Vol-2808/Paper\_29.pdf}

\bibitem{abs-2107-09614}
C.~Birchler, S.~Khatiri, P.~Derakhshanfar, S.~Panichella, A.~Panichella, Single
  and multi-objective test cases prioritization for self-driving cars in
  virtual environments, ACM Transactions on Software Engineering and
  Methodology (TOSEM) (2022).

\bibitem{SmithR21}
S.~C. Smith, S.~Ramamoorthy,
  \href{https://doi.org/10.1109/IROS51168.2021.9636336}{Attainment regions in
  feature-parameter space for high-level debugging in autonomous robots}, in:
  {IEEE/RSJ} International Conference on Intelligent Robots and Systems, {IROS}
  2021, Prague, Czech Republic, September 27 - Oct. 1, 2021, {IEEE}, 2021, pp.
  6546--6551.
\newblock \href {https://doi.org/10.1109/IROS51168.2021.9636336}
  {\path{doi:10.1109/IROS51168.2021.9636336}}.
\newline\urlprefix\url{https://doi.org/10.1109/IROS51168.2021.9636336}

\bibitem{RoyHACC21}
D.~Roy, C.~Hobbs, J.~H. Anderson, M.~Caccamo, S.~Chakraborty,
  \href{https://doi.org/10.23919/DATE51398.2021.9474012}{Timing debugging for
  cyber-physical systems}, in: Design, Automation {\&} Test in Europe
  Conference {\&} Exhibition, {DATE} 2021, Grenoble, France, February 1-5,
  2021, {IEEE}, 2021, pp. 1893--1898.
\newblock \href {https://doi.org/10.23919/DATE51398.2021.9474012}
  {\path{doi:10.23919/DATE51398.2021.9474012}}.
\newline\urlprefix\url{https://doi.org/10.23919/DATE51398.2021.9474012}

\bibitem{afzal2021simulation}
A.~Afzal, D.~S. Katz, C.~Le~Goues, C.~S. Timperley, Simulation for robotics
  test automation: Developer perspectives, in: 2021 14th IEEE Conference on
  Software Testing, Verification and Validation (ICST), IEEE, 2021, pp.
  263--274.

\bibitem{afzal2018crashing}
C.~S. Timperley, A.~Afzal, D.~S. Katz, J.~M. Hernandez, C.~L. Goues,
  \href{http://doi.ieeecomputersociety.org/10.1109/ICST.2018.00040}{Crashing
  simulated planes is cheap: Can simulation detect robotics bugs early?}, in:
  11th {IEEE} International Conference on Software Testing, Verification and
  Validation, {ICST} 2018, V{\"{a}}ster{\aa}s, Sweden, April 9-13, 2018, {IEEE}
  Computer Society, 2018, pp. 331--342.
\newblock \href {https://doi.org/10.1109/ICST.2018.00040}
  {\path{doi:10.1109/ICST.2018.00040}}.
\newline\urlprefix\url{http://doi.ieeecomputersociety.org/10.1109/ICST.2018.00040}

\bibitem{wang2021exploratory}
D.~Wang, S.~Li, G.~Xiao, Y.~Liu, Y.~Sui, An exploratory study of autopilot
  software bugs in unmanned aerial vehicles, in: Proceedings of the 29th ACM
  Joint Meeting on European Software Engineering Conference and Symposium on
  the Foundations of Software Engineering, 2021, pp. 20--31.

\bibitem{Gambi2019Police}
A.~Gambi, T.~Huynh, G.~Fraser,
  \href{https://doi.org/10.1145/3338906.3338942}{Generating effective test
  cases for self-driving cars from police reports}, in: Joint Meeting on
  European Software Engineering Conference and Symposium on the Foundations of
  Software Engineering, {ACM} Press, 2019.
\newblock \href {https://doi.org/10.1145/3338906.3338942}
  {\path{doi:10.1145/3338906.3338942}}.
\newline\urlprefix\url{https://doi.org/10.1145/3338906.3338942}

\bibitem{DosovitskiyRCLK17}
A.~Dosovitskiy, G.~Ros, F.~Codevilla, A.~M. L{\'{o}}pez, V.~Koltun,
  \href{http://proceedings.mlr.press/v78/dosovitskiy17a.html}{{CARLA:} an open
  urban driving simulator}, in: Conference on Robot Learning, Vol.~78 of
  Machine Learning Research, 2017, pp. 1--16.
\newline\urlprefix\url{http://proceedings.mlr.press/v78/dosovitskiy17a.html}

\bibitem{Gambi2019}
A.~Gambi, M.~Mueller, G.~Fraser,
  \href{https://doi.org/10.1109/icse-companion.2019.00030}{{AsFault}: Testing
  self-driving car software using search-based procedural content generation},
  in: 2019 {IEEE}/{ACM} 41st International Conference on Software Engineering:
  Companion Proceedings ({ICSE}-Companion), {IEEE}, 2019.
\newblock \href {https://doi.org/10.1109/icse-companion.2019.00030}
  {\path{doi:10.1109/icse-companion.2019.00030}}.
\newline\urlprefix\url{https://doi.org/10.1109/icse-companion.2019.00030}

\bibitem{abdessalem2018testing}
R.~B. Abdessalem, S.~Nejati, L.~C. Briand, T.~Stifter, Testing vision-based
  control systems using learnable evolutionary algorithms, in: 2018 IEEE/ACM
  40th International Conference on Software Engineering (ICSE), IEEE, 2018, pp.
  1016--1026.

\bibitem{Yoo:2010}
S.~Yoo, M.~Harman, Using hybrid algorithm for {P}areto efficient
  multi-objective test suite minimisation, Journal of Systems and Software
  83~(4) (2010) 689--701.

\bibitem{DBLP:journals/tse/NucciPZL20}
D.~D. Nucci, A.~Panichella, A.~Zaidman, A.~D. Lucia,
  \href{https://doi.org/10.1109/TSE.2018.2868082}{A test case prioritization
  genetic algorithm guided by the hypervolume indicator}, {IEEE} Trans.
  Software Eng. 46~(6) (2020) 674--696.
\newblock \href {https://doi.org/10.1109/TSE.2018.2868082}
  {\path{doi:10.1109/TSE.2018.2868082}}.
\newline\urlprefix\url{https://doi.org/10.1109/TSE.2018.2868082}

\bibitem{SBST2021}
S.~Panichella, A.~Gambi, F.~Zampetti, V.~Riccio, Sbst tool competition 2021,
  in: International Conference on Software Engineering, Workshops, {ACM}, 2021.

\bibitem{CastellanoCTKZA21}
E.~Castellano, A.~Cetinkaya, C.~H. Thanh, S.~Klikovits, X.~Zhang, P.~Arcaini,
  \href{https://doi.org/10.1109/SBST52555.2021.00016}{Frenetic at the {SBST}
  2021 tool competition}, in: International Workshop on Search-Based Software
  Testing, {IEEE}, 2021, pp. 36--37.
\newblock \href {https://doi.org/10.1109/SBST52555.2021.00016}
  {\path{doi:10.1109/SBST52555.2021.00016}}.
\newline\urlprefix\url{https://doi.org/10.1109/SBST52555.2021.00016}

\bibitem{Caruana06anempirical}
R.~Caruana, A.~Niculescu-mizil, An empirical comparison of supervised learning
  algorithms, in: In Proc. 23 rd Intl. Conf. Machine learning (ICML'06, 2006,
  pp. 161--168.

\bibitem{ref1}
C.~Sammut, G.~I. Webb (Eds.),
  \href{https://doi.org/10.1007/978-0-387-30164-8_493}{Logistic Regression},
  Springer US, Boston, MA, 2010, pp. 631--631.
\newblock \href {https://doi.org/10.1007/978-0-387-30164-8_493}
  {\path{doi:10.1007/978-0-387-30164-8_493}}.
\newline\urlprefix\url{https://doi.org/10.1007/978-0-387-30164-8_493}

\bibitem{TinKamHo1998}
T.~K. Ho, \href{https://doi.org/10.1109/34.709601}{The random subspace method
  for constructing decision forests}, {IEEE} Transactions on Pattern Analysis
  and Machine Intelligence 20~(8) (1998) 832--844.
\newblock \href {https://doi.org/10.1109/34.709601}
  {\path{doi:10.1109/34.709601}}.
\newline\urlprefix\url{https://doi.org/10.1109/34.709601}

\bibitem{ke2017lightgbm}
G.~Ke, Q.~Meng, T.~Finley, T.~Wang, W.~Chen, W.~Ma, Q.~Ye, T.-Y. Liu, Lightgbm:
  A highly efficient gradient boosting decision tree, Advances in neural
  information processing systems 30 (2017).

\bibitem{suthaharan2016support}
S.~Suthaharan, Support vector machine, in: Machine learning models and
  algorithms for big data classification, Springer, 2016, pp. 207--235.

\bibitem{safavian1991survey}
S.~R. Safavian, D.~Landgrebe, A survey of decision tree classifier methodology,
  IEEE transactions on systems, man, and cybernetics 21~(3) (1991) 660--674.

\bibitem{PanichellaSGVCG15}
S.~Panichella, A.~D. Sorbo, E.~Guzman, C.~A. Visaggio, G.~Canfora, H.~C. Gall,
  \href{https://doi.org/10.1109/ICSM.2015.7332474}{How can i improve my app?
  classifying user reviews for software maintenance and evolution}, in:
  International Conference on Software Maintenance and Evolution, {IEEE}, 2015,
  pp. 281--290.
\newblock \href {https://doi.org/10.1109/ICSM.2015.7332474}
  {\path{doi:10.1109/ICSM.2015.7332474}}.
\newline\urlprefix\url{https://doi.org/10.1109/ICSM.2015.7332474}

\bibitem{SorboPASVCG16}
A.~{Di Sorbo}, S.~Panichella, C.~V. Alexandru, J.~Shimagaki, C.~A. Visaggio,
  G.~Canfora, H.~C. Gall, What would users change in my app? {S}ummarizing app
  reviews for recommending software changes, in: Proc. Int'l Symposium on
  Foundations of Software Engineering (FSE), 2016, pp. 499--510.

\bibitem{MDG:2020}
F.~Martinez-Taboada, J.~I. Redondo, Induction of decision trees, PLOS ONE
  (2020).

\bibitem{GerstenbergerV15}
C.~Gerstenberger, D.~Vogel, \href{https://doi.org/10.1007/s10260-015-0315-x}{On
  the efficiency of gini's mean difference}, Stat. Methods Appl. 24~(4) (2015)
  569--596.
\newblock \href {https://doi.org/10.1007/s10260-015-0315-x}
  {\path{doi:10.1007/s10260-015-0315-x}}.
\newline\urlprefix\url{https://doi.org/10.1007/s10260-015-0315-x}

\bibitem{9240701}
A.~Trautsch, S.~Herbold, J.~Grabowski, Static source code metrics and static
  analysis warnings for fine-grained just-in-time defect prediction, in: 2020
  IEEE International Conference on Software Maintenance and Evolution (ICSME),
  2020, pp. 127--138.
\newblock \href {https://doi.org/10.1109/ICSME46990.2020.00022}
  {\path{doi:10.1109/ICSME46990.2020.00022}}.

\bibitem{CanforaLPOPP13}
G.~Canfora, A.~D. Lucia, M.~D. Penta, R.~Oliveto, A.~Panichella, S.~Panichella,
  \href{https://doi.org/10.1109/ICST.2013.38}{Multi-objective cross-project
  defect prediction}, in: Sixth {IEEE} International Conference on Software
  Testing, Verification and Validation, {ICST} 2013, Luxembourg, Luxembourg,
  March 18-22, 2013, {IEEE} Computer Society, 2013, pp. 252--261.
\newblock \href {https://doi.org/10.1109/ICST.2013.38}
  {\path{doi:10.1109/ICST.2013.38}}.
\newline\urlprefix\url{https://doi.org/10.1109/ICST.2013.38}

\bibitem{GranoLPP21}
G.~Grano, C.~Laaber, A.~Panichella, S.~Panichella,
  \href{https://doi.org/10.1109/TSE.2019.2946773}{Testing with fewer resources:
  An adaptive approach to performance-aware test case generation}, {IEEE}
  Trans. Software Eng. 47~(11) (2021) 2332--2347.
\newblock \href {https://doi.org/10.1109/TSE.2019.2946773}
  {\path{doi:10.1109/TSE.2019.2946773}}.
\newline\urlprefix\url{https://doi.org/10.1109/TSE.2019.2946773}

\end{thebibliography}
}

\end{document}